\documentclass[10pt]{bmc_article}

\usepackage{epsfig}
\usepackage{cite} 
\usepackage{url}  
\usepackage{ifthen}  
\usepackage{multicol}   
\usepackage[utf8]{inputenc} 

\usepackage{amsmath,amssymb,amsfonts,bm}
\newboolean{publ}

\newenvironment{bmcformat}{\begin{raggedright}\baselineskip20pt\sloppy\setboolean{publ}{false}}{\end{raggedright}\baselineskip20pt\sloppy}

\begin{document}
\begin{bmcformat}

\title{Exploiting Nonlinear Recurrence and Fractal Scaling Properties for Voice Disorder Detection}

\author{Max A Little\correspondingauthor$^{1,2,3}$%
       \email{Max A Little\correspondingauthor - littlem@ieee.org}%
      \and
         Patrick E McSharry$^1$%
         \email{Patrick E McSharry - mcsharry@robots.ox.ac.uk}
      \and
         Stephen J Roberts$^2$%
         \email{Stephen J Roberts - sjrob@robots.ox.ac.uk}
      \and
         Declan AE Costello$^4$%
         \email{Declan AE Costello - declancostello@doctors.org.uk}
       and
         Irene M Moroz$^3$%
         \email{Irene M Moroz - moroz@maths.ox.ac.uk}%
      }

\address{%
    \iid(1)Systems Analysis, Modelling and Prediction Group,%
        Department of Engineering Science, University of Oxford, Parks Road, Oxford OX1 3PJ, UK\\
    \iid(2)Pattern Analysis Research Group,%
        Department of Engineering Science, University of Oxford, Parks Road, Oxford OX1 3PJ, UK\\
    \iid(3)Applied Dynamical Systems Research Group,%
        Oxford Centre for Industrial and Applied Mathematics, Mathematics Institute,%
        University of Oxford, Oxford OX1 3JP, UK\\
    \iid(4)Milton Keynes General Hospital, Standing Way, Eaglestone, Milton Keynes, Bucks MK6 5LD, UK}%

\maketitle

\begin{abstract}
        \paragraph*{Background:} Voice disorders affect patients profoundly, and acoustic tools can
        potentially measure voice function objectively. Disordered sustained vowels exhibit wide-ranging
        phenomena, from nearly periodic to highly complex, aperiodic vibrations, and increased ``breathiness''.
        Modelling and surrogate data studies have shown significant nonlinear and non-Gaussian random properties
        in these sounds. Nonetheless, existing tools are limited to analysing voices displaying near periodicity,
        and do not account for this inherent biophysical nonlinearity and non-Gaussian randomness, often using
        linear signal processing methods insensitive to these properties. They do not directly measure the two
        main biophysical symptoms of disorder: complex nonlinear aperiodicity, and turbulent, aeroacoustic,
        non-Gaussian randomness. Often these tools cannot be applied to more severe disordered voices,
        limiting their clinical usefulness.

        \paragraph*{Methods:} This paper introduces two new tools to speech analysis: recurrence and fractal scaling,
        which overcome the range limitations of existing tools by addressing directly these two symptoms of disorder,
        together reproducing a ``hoarseness'' diagram. A simple bootstrapped classifier then uses these two features to
        distinguish normal from disordered voices.

        \paragraph*{Results:} On a large database of subjects with a wide variety of voice disorders,
        these new techniques can distinguish normal from disordered cases, using quadratic discriminant analysis,
        to overall correct classification performance of $91.8 \pm 2.0\%$. The true positive classification performance
        is $95.4 \pm 3.2\%$, and the true negative performance is $91.5 \pm 2.3\%$ ($95\%$ confidence).
        This is shown to outperform all combinations of the most popular classical tools.

        \paragraph*{Conclusions:} Given the very large number of arbitrary parameters and computational complexity
        of existing techniques, these new techniques are far simpler and yet achieve clinically useful classification
        performance using only a basic classification technique. They do so by exploiting the inherent nonlinearity
        and turbulent randomness in disordered voice signals. They are widely applicable to the whole range of disordered
        voice phenomena by design. These new measures could therefore be used for a variety of practical clinical purposes.
\end{abstract}

\ifthenelse{\boolean{publ}}{\begin{multicols}{2}}{}

\section*{Background}
Voice disorders arise due to physiological disease or psychological disorder, accident, misuse of the voice, or surgery affecting the vocal folds and have a profound
impact on the lives of patients. This effect is even more extreme when the patients are professional voice users, such as singers, actors, radio and television presenters,
for example. Commonly used by speech clinicians, such as surgeons and speech therapists, are acoustic tools, recording changes in acoustic pressure at the lips or
inside the vocal tract. These tools \cite{baken}, amongst others, can provide potentially objective measures of voice function. Although acoustic examination is
only one tool in the complete assessment of voice function, such objective measurement has many practical uses in clinical settings, augmenting the subjective
judgement of voice function by clinicians. These measures find uses, for example, in the evaluation of surgical procedures, therapy, differential diagnosis and
screening \cite{baken, carding}, and often augment subjective voice quality measurements, for example the GRB (Grade, Roughness and Breathiness) scale.\cite{dejonckere}
These objective measures can be used to portray a ``hoarseness'' diagram for clinical applications \cite{michaelis}, and there also exists
a variety of techniques for automatically screening for voice disorders using these measures \cite{boyanov, godino-04, alonso}.

Phenomenologically, normal and disordered sustained vowel speech sounds exhibit a large range of behaviour. This includes {\it nearly periodic} or regular
vibration, {\it aperiodic} or irregular vibration and sounds with no apparent vibration at all. All can be accompanied by varying degrees
of noise which can be described as ``breathiness''. Voice disorders therefore commonly exhibit two characteristic phenomena: increased vibrational
aperiodicity and increased breathiness compared to normal voices \cite{michaelis}.

In order to better characterise the vibrational aperiodicity aspects of voice disorders, Titze \cite{titze-95}
introduced a typology (extended with subtypes \cite{baken}). Type I sounds are those that are nearly periodic: coming close to perfect periodicity.
Type II are those that show qualitative dynamical changes and/or modulating frequencies or subharmonics. The third class, Type III are those sounds that appear
to have no periodic pattern at all. They have no single, obvious or dominant period and can described as {\it aperiodic}. Normal voices can usually be classed as
Type I and sometimes Type II, whereas voice disorders commonly lead to all three types of sounds. This is because
voice disorders often cause the breakdown of stable periodicity in voice production. The breathiness aspect of disordered voices is often described as
dominating high-frequency noise. Although this original typology covered sounds of only apparently deterministic origin, a very large proportion
of voice signals seen in the clinic are so noisy as to be better considered stochastic rather than deterministic \cite{carding}.
Methods that are based upon theories of purely deterministic nonlinear dynamical systems, although they can be appropriate for sounds of deterministic
origin covered by the original typology, cannot in principle be applied to such noise-dominated sounds, that is, to sounds that would be better modelled
as stochastic processes rather than deterministic. This makes it impossible to characterise the full range of signals encountered in the clinic.
For these reasons, in this paper, when we refer to Type III sounds we include random, noise-like sounds (which, in keeping with original typology,
have no apparent periodic structure, by virtue of their randomness).

There exists a very large number of approaches to the acoustic measurement of voice function. The most popular
of these are the {\it perturbation} measures {\it jitter} and {\it shimmer} and variants, and {\it harmonics-to-noise ratios (HNR)} \cite{baken, michaelis}.
The effect of a variety of voice disorders on these measures has been tested under both experimental and clinical conditions \cite{schoentgen, michaelis}, showing
that different measures respond to different disorders in different ways \cite{hirano-88}. For example, in some disorders, jitter will increase with
severity of the disorder, and for other disorders jitter is unaffected. Thus, although these measures can have value in
measuring certain limited aspects of voice disorders such as speech pitch variability, there is no simple relationship
between the extent or severity of voice disorder
and these measures \cite{hirano-88, michaelis}. Therefore, they cannot be used to directly quantify the two main biophysical
symptoms of voice disorders: increasingly severe aperiodicity and breath noise, a quantification required to
differentiate normal from disordered voices.

Another limitation of existing measures is that they are only properly applicable when near periodicity holds:
in Titze's typology only Type I sounds satisfy this property \cite{baken}. The behaviour of the algorithms for other sound types is not known theoretically and
limited only to experimental results and informal arguments \cite{michaelis}. The source of this limitation is that they make extensive use of extraction of the {\it pitch
period}, or {\it fundamental frequency} (defined as the inverse of the pitch period) from the acoustic signal \cite{baken}. Popular pitch period extraction
techniques include zero-crossing detection, peak-picking and waveform matching \cite{baken}. The concept of pitch period is only valid for Type I sounds,
therefore, application of these methods based upon periodicity analysis, to any other type of sound is problematic \cite{godino-04}.
Type II and III have therefore received much less attention in the literature \cite{titze-95}, such that there exist few methods for characterising these types,
despite the fact that they exist in great abundance in clinical settings. This precludes the proper use of these tools on a large number of disordered
voice cases, limiting the reliability of the analysis, since in fact some algorithms will not produce any results at all for Type II and III sounds, even though they
contain valuable clinical information \cite{carding}. Another reason for the limitations of these methods is that they are based upon classical linear signal processing
methods that are insensitive to the inherent biophysical nonlinearity and non-Gaussianity in speech \cite{baken}. These linear methods include autocorrelation, the discrete
Fourier transform, linear prediction analysis and cepstral processing \cite{quatieri}.

Since standardised, reliable and reproducible results from acoustic measures of voice function are required for clinical applications, these limitations of perturbation
methods are problematic in clinical practice \cite{carding}. It is clear that there is a clinical need for reliable tools that can characterise all types of disordered voice sounds for
a variety of clinical applications, regardless of whether they satisfy the requirements of near periodicity, or contain significant nonlinearity,
randomness or non-Gaussianity \cite{titze-95}.

Furthermore, analysis techniques are complicated by the use of any arbitrary algorithmic parameters whose choice affects
the analysis method -- changing these parameters can change the analysis results. The values of such parameters must be chosen in order to apply the
algorithms, and to be principled, it is better, when making that choice, to have a theoretical framework to apply which offers specific guidance
on that choice. Often however, no such general theory exists, and therefore these values must be chosen by experimental and empirical evaluation alone \cite{box}.
There exists the danger then that these parameters are highly ``tuned'' to the particular data set used in any one study, limiting the reproducibility of the analysis
on different data sets or clinical settings. It is necessary therefore to reduce the number arbitrary parameters to improve the reproducibility of these measures.

To address these limitations, empirical investigations and theoretical modelling studies have been conducted which have lead to the suggestion that
{\it nonlinear dynamical systems theory} is a candidate for a unified mathematical framework modelling the dynamics seen in all types of disordered
vowel speech sounds \cite{herzel-94, titze-95, baken}. The motivation for the introduction of a more general model than the classical linear model
is the principle of parsimony: the more general model explains more phenomena (more types of speech) with a smaller number of assumptions
than the classical linear model \cite{box, akaike}.

These suggestions have lead to growing interest in applying tools from nonlinear time series analysis to speech signals in order to attempt to
characterise and exploit these nonlinear phenomena \cite{baken, herzel-94}. For normal Type I speech sounds,
fractal dimensions, Lyapunov exponents and bispectral methods have all been applied, also finding evidence to support the existence
of nonlinearity \cite{maragos-99, banbrook}. Extracting dynamical structure using local linear predictors, neural networks and regularised radial basis functions
have all been used, with varying reported degrees of success. Algorithms for calculating the correlation dimension have been
applied, which was successful in separating normal from disordered subjects \cite{zhang-05}.
Correlation dimension and second-order dynamical entropy measures showed statistically
significant changes before and after surgical intervention for vocal fold polyps \cite{zhang-mcgilligan-04}, and
Lyapunov exponents for disordered voices were found to consistently higher than those for healthy voices \cite{giovanni}.
It was also found that jitter and shimmer measurements were less reliable than correlation dimension analysis on Type I and unable to characterise Type II
and (non-random) Type III sounds \cite{zhang-jasa-05}. Mixed results were found for fractal dimension analysis of sounds from patients
with neurological disorders, for both acoustic and electroglottographic signals \cite{hertrich}. Instantaneous nonlinear AM and FM formant modulations were shown
effective at detecting muscle tension dysphonias \cite{hansen}. For the automated acoustic screening of voice disorders, higher-order statistics lead to
improved normal/disordered classification performance when combined with several standard perturbation methods \cite{alonso}.

In order to categorise individual voice signals into classes from the original typology (excluding severely turbulent, noise-like sounds),
recent studies have found that by applying correlation dimension measurements to signals of these types, it was possible, over a sample of 122 stable vowel phonations,
to detect a statistically significant difference between the three different classes \cite{zhang-type03, jiang-jvoice06}. This provides further evidence in favour of the
appropriateness of nonlinear signal analysis techniques for the analysis of disordered voices.

These studies show that nonlinear time series methods can be valuable tools for the analysis of voice disorders, in that they can analyse a much broader
range of speech sounds than perturbation measures, and in some cases are found to be more reliable under conditions of high noise.
However, very noisy, breathy signals have so far received little attention from nonlinear time series analysis approaches,
despite these promising successes. Common approaches such as
correlation dimension, Lyapunov exponent calculation and predictor construction require that the scaling properties of the embedded attractor are not
destroyed by noise, and for thus very noisy, breathy signals, there is the possibility that such nonlinear approaches may fail. There are also numerical, theoretical
and algorithmic problems associated with the calculation of nonlinear measures such as Lyapunov exponents or correlation dimensions for real speech signals, casting doubt
over the reliability of such tools \cite{kantz, behrman-97, hertrich, little-icassp}. For example, correlation dimension analysis shows high sensitivity to the variance
of signals in general, and it is therefore necessary to check that changes in correlation dimension are not due simply to changes in variance \cite{mcsharry-03}.
Therefore, as with classical perturbation methods, current nonlinear approaches cannot yet directly measure the two most important biophysical aspects of voice
disorder.

A limitation of deterministic nonlinear time series analysis techniques for random, very noisy signals is that the implicit,
deterministic, nonlinear dynamical model, which is ordinarily assumed to represent the nonlinear oscillations of the vocal
folds \cite{herzel-95} is no longer appropriate. This is because randomness is an inherent part of the biophysics of speech
production \cite{jiang-02, krane}. Thus there is a need to expand
the nonlinear dynamical systems framework to include a random component, such that random voice signals and breath noise
can also be fully characterised within the same, unified framework.

This paper therefore introduces a new, framework model of speech production that splits the dynamics into both deterministic nonlinear {\it and} stochastic
components. The output of this model can then be analysed using methods that are able to characterise both nonlinearity and randomness.
The deterministic component of the model can exhibit both periodic and aperiodic dynamics. It is proposed to characterise this component using
{\it recurrence analysis}. The stochastic components can exhibit statistical time dependence or autocorrelation, which can be analysed effectively using
{\it fractal scaling analysis}. This paper reports the replication of the ``hoarseness'' diagram \cite{michaelis} illustrating the extent of voice disorder, and
demonstrates, using a simple quadratic classifier, how these new measures may be used to screen normal from disordered voices from a large,
widely-used database of patients. It also demonstrates that these new measures achieve superior classification performance
overall when compared to existing, classical perturbation measures, and the derived irregularity and noise measures of Michaelis \cite{michaelis}.

\section*{Methods}
In this section we first discuss in detail the evidence that supports the development of a new stochastic/deterministic model of speech production.

The classical linear theory of speech production brings together the well-developed subjects of classical linear signal processing and
linear acoustics (where any nonlinearities in the constitutive relationships between dynamic variables of the fluid are considered small enough to be negligible)
to process and analyse speech time series \cite{quatieri}. The biophysical, acoustic assumption is that the vocal tract can be modelled
as a linear resonator driven by the vibrating vocal folds that are the source of excitation energy \cite{quatieri}.
However, extensive modelling studies, experimental investigations and analysis of voice signals have shown that dynamical nonlinearity and
randomness are factors that should not be ignored in modelling speech production.

For the vocal folds, there are two basic, relevant model components to consider. The first is the vocal fold tissue, and the second is the air flowing through that structure.
The vocal folds, during phonation, act as a vibrating valve, disrupting the constant airflow coming from the lungs and
forming it into regular puffs of air. In general the governing equations are those of fluid dynamics coupled with the elastodynamics of a deformable solid.
In one approach to solving the problem, the airflow is modelled as a modified quasi-one-dimensional Euler system which is coupled to the vocal fold flow rate, and the vocal folds are
modelled by a lumped two mass system \cite{lamar}. A widely used and simplified model is the two-mass model in Ishizaka \cite{ishizaka}, further simplified in
asymmetric \cite{herzel-95, steinecke-95} and symmetric versions \cite{jiang-01}. These models demonstrate a plausible physical mechanism for phonation as nonlinear
oscillation: dynamical forcing from the lungs supplies the energy needed to overcome dissipation
in the vocal fold tissue and vocal tract air. The vocal folds themselves are modelled as elastic tissue with nonlinear
stress-strain relationship.

Classical nonlinear vocal fold models coupled with linear acoustic vocal tract resonators appear to account for the major part of the mechanisms of audible speech,
but from these mechanisms an important component is missing: that of {\it aspiration}, or ``breath'' noise \cite{dixit}.
Such noise is produced when the air is forced through a narrow constriction at sufficiently high speeds that ``turbulent'' airflow is generated, which in turn
produces noise-like pressure fluctuations \cite{howe}. Aspiration noise is an unavoidable, involuntary consequence of airflow from the lungs being forced through the vocal organs,
and can be clearly heard in vowel phonation \cite{sinder}. Certain voice pathologies are accompanied by a significant increase
in such aspiration noise, which is perceived as increased ``breathiness'' in speech \cite{michaelis}. This noise is therefore an important part of sound generation in speech.

One significant deficiency in the classical linear acoustic models is due to the assumptions about fluid flow upon which their construction is based \cite{mclaughlin-06}.
These classical linear models make very many simplifying assumptions about the airflow in the vocal organs, for example, that the
{\it acoustic limit} \cite{ockendon} holds in which the fluid is nearly in a state of uniform motion. Similarly, the {\it Bernoulli's equation} assuming energy conservation along streamlines,
upon which many classical vocal fold models rely, applies only if the fluid is assumed inviscid and irrotational \cite{acheson, kinsler}.
The important point for this study is that these classical assumptions forbid
the development of complicated, ``turbulent'' fluid flow motion, in which the flow follows convoluted paths of rapidly varying velocity, with eddies and other irregularities
at all spatial scales \cite{falconer}. This breakdown of laminar flow occurs at high {\it Reynolds number}, and for the relevant length scales of a few centimetres in the vocal tract and for
subsonic air flow speeds typical of speech \cite{sinder}, this number is very large (of order $10^5$), indicating that airflow in the vocal tract can be
expected to be turbulent. Under certain assumptions, turbulent structures, and vortices in particular (fluid particles that have rotational motion), can be shown to be a source
of aeroacoustic sound \cite{howe}. Turbulence is a highly complex dynamical phenomenon and any point measurement such as acoustic pressure taken
in the fluid will lose most of the information about the dynamics of the fluid. Consequently, even if turbulence has
a purely deterministic origin, it is reasonable to model any one single dynamical variable measured at a point in space as a random process \cite{falconer}.

There are two broad classes of physically plausible mathematical models of the effects of aeroacoustic noise
generation in speech. The first involves solving numerically the full partial differential equations of gas dynamics (e.g. the Navier-Stokes equations), and the second uses the
theory of {\it vortex sound} \cite{howe}. For example, the study of Zhao \cite{zhao-02} focused on the production
of aspiration noise generated by vortex shedding at the top of the vocal folds, simulated over a full vocal fold cycle.
This study demonstrates that the computed sound radiation due to vorticity contains significant high frequency fluctuations when the folds
are fully open and beginning to close. On the basis of these results, it can be expected that if the folds do not close completely during a cycle
(which is observed in cases of more ``breathy'' speech), the amplitude of high frequency noise will increase.

The second class of models, which makes use of {\it Lighthill's acoustic analogy} \cite{howe}, are based around the theory of vortex sound generated in a cylindrical duct,\cite{howe}
where, essentially, each vortex shed at an upstream constriction acts as a source term for the acoustic wave equation in the duct, as the vortex is convected along with the steady
part of the airflow. The resulting source term depends upon not only the attributes of the vortex itself, but also upon the motion of the vortex through
the streamlines of the flow \cite{krane, howe}. One model that uses this approach involves the numerical simulation of two overall components:
the mean steady flow field and the acoustic wave propagation in the vocal tract \cite{sinder}. Vortices are assumed to be shed at random intervals at constrictions at particular
locations in the vocal tract, for example, at the vocal folds or between the roof of the mouth and the tongue. Each vortex contributes to the acoustic source term at each spatial grid
point.
Here, an important observation is that simulated pressure signals from this model are stochastic processes \cite{grimmett}, i.e. a sequence of random variables.
It is also noticeable from the spectra of simulated pressure signals that although the signals are stochastic, they exhibit significant non-zero autocorrelation
since the spectral magnitudes are not entirely constant, leading to statistical self-similarity in these signals \cite{falconer, maragos-99}.

Other potential sources of biophysical fluctuation include pulsatile blood flow, muscle twitch and other variability and randomness
in the neurological and biomechanical systems of the larynx \cite{jiang-02}.

The typical nonlinear dynamical behaviour of models of the vocal folds, such as period-doubling (subharmonics), bifurcations \cite{herzel-95}, and transitions to irregular vibration \cite{
jiang-01}
have been observed in experiments with excised larynxes, a finding that helps to support the modelling hypothesis that speech is an inherently nonlinear phenomena \cite{berry-96,
herzel-95}.
Similarly, models of turbulent, random aeroacoustic aspiration noise have been validated against real turbulent airflow induced sound generated in acoustic duct experiments \cite{sinder}.
Such studies show that the models of vortex shedding at random intervals are plausible accounts for the dynamical origins of breath noise in phonation.

Complementing modelling and experimental studies, the final source of evidence for nonlinearity and randomness in speech signals comes from studies of voice pressure signals.
Using surrogate data analysis, it has been shown that nonlinearity and/or non-Gaussianity is an important feature of Type I sounds \cite{little-jasa, tokuda-01, tokuda-96}.
Nonlinear bifurcations have been observed in excised larynx experiments \cite{berry-96}, and period-doubling bifurcations and chaos-like features have been observed in
signals from patients with organic and functional dysphonias \cite{herzel-94}. Aspiration noise has been observed and measured as a source of randomness
in voiced phonation, in both normal \cite{jackson} and disordered speech sounds \cite{baken, michaelis}. The fractal, self-similarity properties of aspiration noise as a turbulent sound
source
have also been observed and quantified in normal \cite{maragos-99} and disordered speech \cite{little-icassp}.

Taken as a whole, these modelling, simulation, validation and signal analysis studies suggest that during voiced phonation there will be a combination of both deterministic and stochastic
elements,
the deterministic component attributable to the nonlinear movements of the vocal fold tissue and bulk of the air in the vocal tract, and the stochastic component the high frequency
aeroacoustic pressure fluctuations caused by vortex shedding at the top of the vocal folds, whose frequency and intensity is modulated by the bulk air movement (and other sources of
biophysical fluctuation). During voiced phonation, the deterministic oscillations will dominate in amplitude over the noise component which will show high frequency fluctuations around
this oscillation.
During unvoiced or breathy pathological phonation, the turbulent noise component will dominate over any deterministic motion.

In order to capture all these effects in one unified model, we introduce a continuous time, two component dynamical model that is taken to generate the measured
speech signal. The state of the system at time $t \in \mathbb{R}$ is represented by the vector ${\bf u}(t)$ of size $d$.
Then the equation of motion that generates the speech signal is the following vector stochastic differential equation,
commonly known as a {\it Langevin equation} \cite{kantz}.
\begin{equation}
\label{stoch_model}
{\bf \dot u}\left( t \right) = {\bf f}\left( {{\bf u}\left( t \right)} \right) + {\bf \varepsilon }\left( t \right),
\end{equation}
where ${\bf \varepsilon}(t)$ is a vector of stochastic forcing terms. It is not necessary to assume that these fluctuations are
independent and identically distributed (i.i.d.). The function ${\bf f} : \mathbb{R}^d \to \mathbb{R}^d$ is unknown
and represents the deterministic part of the dynamics. Given an initial condition vector ${\bf u}(0)$ then a solution that
satisfies equation \eqref{stoch_model} is called a {\it trajectory}. Ensembles of trajectories can be shown to satisfy the
properties of a stochastic process with finite memory (a higher-order Markov chain) \cite{kantz}.

Of importance to both deterministic and stochastic systems is the notion of recurrence in state space. Recurrent trajectories are those
that return to a given region of state space \cite{altmann}. Recurrence time statistics provide useful information about the properties
of both purely deterministic dynamical systems and stochastic systems \cite{altmann}. Recurrence analysis forms the basis of the
method of recurrence plots in nonlinear time series analysis \cite{kantz}.

In the context of the model \eqref{stoch_model}, state-space recurrence is defined as:
\begin{equation}
\label{recurrence}
{\bf u}(t) \subset B({\bf u}(t + \delta t), r),
\end{equation}
where $B({\bf u}, r)$ is a closed ball of small radius $r > 0$ around the point ${\bf u}$ in state-space, and
${\bf u}(t) \not\subset B({\bf u}(t + s), r)$ for $0 < s < \delta t$. Each different $t \in \mathbb{R}$ may be associated
with a different $\delta t$, called the {\it recurrence time}. We will define {\it aperiodic} as recurrent in the
sense of \eqref{recurrence} but not {\it periodic}. Periodicity is recovered from the definition of recurrence in the special case
when $r = 0$ and $\delta t$ is the same for all $t$, so that the system vector ${\bf u}(t)$ assumes the same value after a time interval of $\delta t$:
\begin{equation}
{\bf u}(t) = {\bf u}(t + \delta t),
\end{equation}
for all $t \in \mathbb{R}$. Then $\delta t$ is the {\it period} of the system.
Therefore, although these concepts of periodicity and aperiodicity are mutually exclusive,
both are special cases of the more general concept of recurrence. The requirement of periodicity is central to many linear signal processing methods
(the basis of the Fourier transform, for example), but aperiodicity is a common feature of many voice disorders. It can be seen therefore that in
order to characterise the inherent irregularity of disordered speech sounds, we require more general processing techniques that can directly account
for such departures from strict periodicity.

Using this analysis, nearly periodic speech sounds of Type I can be described as recurrent for some small $r > 0$, with $\delta t$ nearly the same
for each $t$. Type II sounds are more irregular than Type I, and for the same $r$, the $\delta t$ will assume a wider range of values than for Type I.
Similarly, Type III sounds which are highly irregular and aperiodic, will have a large range of values of $\delta t$ again for the same $r$.

Similarly, of importance in the analysis of stochastic processes is scaling analysis \cite{kantz}. Stochastic time series in which the individual measurements in time are
not entirely independent of earlier time instants are often {\it self-similar}, that is, when a sub-section of the time series is scaled up by a certain factor, it has
geometrical, approximate or statistical similarity to the whole time series \cite{kantz}. Such self-similarity is a property of {\it fractals}.\cite{falconer} The tools of dimension
measurement and scaling analysis may be used to characterise the self-similarity in signals such as speech.
As discussed above, theoretical models of aeroacoustic turbulent sound generation in speech
predict randomly-timed impulses convolved with an individual impulse response for each vortex that induces {\it autocorrelated} random noise sequences \cite{krane},
so that turbulent speech signals at one time instant are not independent of earlier time instants.

It has also been found experimentally that changes in the statistical time dependence properties of turbulent noise in speech, as measured by a particular fractal
dimension of the graph of the speech signal, are capable of distinguishing classes of phonemes from each other \cite{maragos-99}. As introduced above, disordered
voices are often accompanied by increased ``breathiness'', due in part to the inability of the vocal folds to close properly, so that air escapes
through the partial constriction of the vocal folds creating increased turbulence in the airflow \cite{krane}. Thus scaling analysis (and more general graph dimension
measures) could be useful for characterising vocal fold disorders.

Recent experiments have shown that the use of recurrence analysis coupled with scaling analysis can distinguish healthy from disordered
voices on a large database of recordings with high accuracy \cite{little-icassp}. These techniques are computationally simple and furthermore
substantially reduce the number of arbitrary algorithmic parameters required, compared to existing classical measures, thus leading to increased reproducibility and reliability.

\subsection*{Time-Delay State-Space Recurrence Analysis}
In order to devise a practical method for applying the concept of recurrence defined earlier and measuring the extent of aperiodicity in a speech signal, we can make
use of {\it time-delay embedding} \cite{kantz}. Measurements of the output of the system \eqref{stoch_model} are assumed to constitute the acoustic signal, $s_n$:
\begin{equation}
s_n  = h\left( {{\bf u}\left( {n\Delta t} \right)} \right),
\end{equation}
where the measurement function $h : \mathbb{R}^d \to \mathbb{R}$ projects the state vector ${\bf u}(t)$ onto
the discrete-time signal at time instances $n \Delta t$ where $\Delta t$ is the sampling time, and $n \in \mathbb{Z}$ is the time index.

From these sampled measurements, we then construct of an $m$-dimensional {\it time delay embedding vector}:
\begin{equation}
\label{delay_map}
{\bf s}_n  = \left[ {s_n ,s_{n - \tau } , \ldots s_{n - \left( {m - 1} \right)\tau } } \right]^T .
\end{equation}
Here $\tau$ is the {\it embedding time delay} and $m$ is the {\it embedding dimension}.

We will make use of the approach introduced in Ragwitz \cite{ragwitz} to {\it optimise}
the embedding parameters $m$ and $\tau$ such that the recurrence analysis produces results that are close as possible to known,
analytically derived results upon calibration with known signals. (We use this as an alternative to common techniques
for finding embedding parameters, such as false-nearest neighbours, which explicitly require purely deterministic dynamics \cite{kantz, ragwitz}).
Note that, under this approach, for very noisy signals, we will not always resolve all signals without self-intersections.
However, in the context of this study, achieving a completely non-intersecting embedding is not necessary.
For very high-dimensional deterministic or stochastic systems, any reconstruction with self-intersections due to
insufficiently high embedding dimension can be considered as a different stochastic system in the reconstructed state-space.
We can then analyze the stochastic recurrence of this reconstructed system. This recurrence, albeit different from the
recurrence properties in the original system, is often sufficient to characterize the noisy end of the scale of
periodicity and aperiodicity.

Figure 1 shows the signals $s_n$ for one normal and one disordered voice example (Kay Elemetrics
Disordered Voice Database). Figure 2 shows the result of applying the above embedding procedure for the same speech signals.

In order to investigate practically the recurrence time statistics of the speech signal, we can make use of
the {\it method of close returns} \cite{lathrop-89}, originally designed for studying recurrence in deterministic, chaotic dynamics.
Here we adopt this method to study stochastic dynamics as well as deterministic dynamics.
In this method, a small, closed ball $B({\bf s}_{n_{0}},r)$ of radius $r > 0$ is placed around the embedded data point ${\bf s}_{n_0}$. Then the
trajectory is followed forward in time ${\bf s}_{n_{0}+1}, {\bf s}_{n_{0}+2} \ldots$ until it has left this ball, i.e.
until $\left\| {{\bf s}_{n_0 }  - {\bf s}_{n_0  + j} } \right\| > r$ for some $j > 0$, where $\left\|.\right\|$ is the Euclidean distance.
Subsequently, the time $n_1$ at which the trajectory first returns to this same ball is recorded (i.e. the first time when
$\left\| {{\bf s}_{n_0 }  - {\bf s}_{n_1 } } \right\| \le r$), and the difference of these two times is the (discrete) {\it recurrence time} $T = n_1 - n_0$.
This procedure is repeated for all the embedded data points ${\bf s}_n$, forming a histogram of recurrence times $R(T)$.
This histogram is normalised to give the {\it recurrence time probability density}:
\begin{equation}
\label{rtpd}
P( T ) = \frac{R(T)}{\sum_{i = 1}^{T_{\max}} R(i)},
\end{equation}
where $T_{\max}$ is the maximum recurrence time found in the embedded state space. The choice of $r$ is critical to capture the properties
of interest to this study. For example, if the trajectory is a nearly periodic (Type I), we require that $r$ is large enough to capture all the recurrences,
but not too large to find recurrences that are due to spurious intersections of $B({\bf u}, r)$ with other parts of the trajectory, violating the
conditions for proper recurrence. The appropriate choice of embedding delay $\tau$ has a role to play: selecting $\tau$ too small means
that any trajectory lies close to a diagonal in the reconstructed state-space, potentially causing spurious recurrences. Thus $\tau$
must be chosen large enough to avoid spurious recurrences. Similarly, if $\tau$ is too large then the points in the
embedded state-space fill a large cloud where recurrences will be difficult to find without
using an inappropriately large value of $r$. There will be an optimum value of $\tau$ which traditionally is set
with reference to autocorrelation or time-delayed mutual information estimates, for more details see \cite{kantz}.

In order to understand the behaviour of this algorithm (partly for optimising the embedding parameters), we consider two extreme forms that the
density \eqref{rtpd} may assume. The first is the ideal limiting case in which the recurrence distance
$r$ tends to zero for a periodic trajectory. The recurrence time probability density is:
\begin{equation}
\label{theory_periodic}\
P( T ) = \left\{
\begin{array}{l}
 1\quad {\rm if}\;T = k \\
 0\quad {\rm otherwise} \\
 \end{array} \right .,
\end{equation}
where $k$ is the period of the trajectory. In the second extreme case we consider a purely random,
uniform i.i.d. signal which is normalised to the range $\left[ -1, 1\right]$. The recurrence probability density is approximately uniform:
\begin{equation}
\label{theory_uniformiid}
P(T) \approx \frac{1}{T_{\rm max}}.
\end{equation}

Proofs for the results in equations \eqref{theory_periodic} and \eqref{theory_uniformiid} are given in the Appendix.

We can then optimise $m$, $\tau$ and $r$ such that for a synthetic signal of perfect periodicity, $P(T)$ is determined using the close returns method
such that it is as close as possible to the theoretical expression \eqref{theory_periodic}. This optimisation is carried out by a straightforward
systematic search of values of these parameters $m = 2,3 \ldots 10$, $\tau = 2, 3 \ldots 50$, and for $r = 0.02, 0.04, \ldots 0.5$,
on a perfectly periodic test signal. This search can be considered as a scan for the optimum parameter values through all points on a three-dimensional parameter
cube with $m$, $\tau$ and $r$ as the co-ordinates of that cube.

All disordered voice speech signals will lie somewhere in between the extremes of perfect periodicity and complete randomness. Thus it will
be useful to create, from the recurrence time probability density, a straightforward sliding scale so that normal and disordered voice signals can be ranked
alongside each other. This depends upon a simple
characterisation of the recurrence probability density $P(T)$. One simple measure of any probability density is the {\it entropy}
which measures the average uncertainty in the value of the discrete-valued density $p(i)$, $i = 1, 2 \ldots M$ \cite{kantz}:
\begin{equation}
H = -\sum_{i = 1}^M p(i) \ln p(i),
\end{equation}
with units of {\it nats} (by convention, $0 \ln 0$ is taken to be zero). This measure can then rank disordered voice signals by representing
the {\it uncertainty in the period} of the disordered voice signal in the following way. For perfectly periodic signals the recurrence probability
density entropy (RPDE) is:
\begin{align}
H_{\rm per} = -\sum_{i = 1}^{T_{\rm max}} P(i) \ln P(i) = 0.
\end{align}
since $P(k) = 1$ and the rest are zero. Conversely, for the purely stochastic, uniform i.i.d. case (derived in the appendix), the
uniform density can be taken as a good approximation, so that the RPDE is:
\begin{equation}
H_{\rm iid} = -\sum_{i=1}^{T_{\rm max}} P(i) \ln P(i) = \ln T_{\rm max} ,
\end{equation}
in units of nats. The entropy scale $H$ therefore ranges from $H_{\rm per}$, representing perfectly periodic examples of Type I sounds,
to $H_{\rm iid}$ as the most extreme cases of noise-like Type III sounds. In practice, all sounds will lie somewhere in between these
extremes.

However, the entropy of a probability density is maximum for the uniform density, so that $H_{\rm iid}$ is the maximum value that $H$ can assume.
Thus, in addition to ranking signals on a scale of aperiodicity, we can know precisely the two extremes of that scale.
For different sampling times $\Delta t$ the value $T_{\rm max}$ will change. Therefore, we can normalised the RPDE scale for subsequent usage:
\begin{equation}
H_{\rm norm} = \frac{-\sum_{i=1}^{T_{\rm max}} P(i) \ln P(i) } {H_{\rm iid}},
\end{equation}
a unit less quantity that assumes real values in the range $[0, 1]$.

The method of close returns, upon which this technique is based, was originally designed to characterise deterministic, chaotic systems \cite{lathrop-89}.
In this case, if the chaotic system is ergodic and has evolved past any transients, then the recurrence properties of the system are independent of the
initial conditions and initial time, i.e. they are invariants of the system. Similarly, if the system is purely stochastic and ergodic, then it has a
stationary distribution. Again, after any transient phase, the recurrence properties will be invariant in the above sense. Thus the derived measure
$H$ will also be an invariant of the system. We note that traditional jitter and shimmer measurements do not share this invariance property,
in this sense they do not give a reliable description for chaotic or Type III time series. Often, when stable phonation is initiated in speech,
the vocal folds will take some time to settle into a pattern of stable periodic or chaotic oscillation. The behaviour of speech signals during
this ``settling time'' is similar to the transient behaviour typically observed in nonlinear dynamical systems.
In this study, we make use of voice signals which are stable phonations, and we discard any of these transient phases. Thus, to a reasonable
approximation $H$ can be considered as an invariant of the dynamics of the speech organs.

Figure 3 shows the result of this recurrence analysis, applied to a synthetic, perfectly periodic signal created by taking a single cycle
from a speech signal and repeating it end-to-end many times. It also shows the analysis applied to a synthesised, uniform, i.i.d. random signal
(on the signal range $[-1, 1]$) after optimising $m$, $\tau$ and $r$ by gridded search. Even though exact results are impossible to obtain due to the
approximation inherent to the algorithm and only finite-length signals, the figure shows that a close match is obtainable between the theoretical, predicted results
and the simulated results.

\subsection*{Detrended Fluctuation Analysis}
In order to investigate the second aspect of disordered voices, that of increased breath noise, we require a practical approach to applying the
scaling ideas introduced above. Detrended fluctuation analysis is one straightforward technique for characterising the self-similarity
of the graph of a signal from a stochastic process \cite{hu}.

It is designed to calculate the {\it scaling exponent} $\alpha$ in nonstationary time series (where the statistics such as mean, variance and autocorrelation
properties might change with time), and has been shown to produce robust results when there are slowly moving trends in the data. These
will naturally include low frequency vibrations \cite{hu}, which are similar in nature to the nonlinear vibrations of the vocal folds described by the
function ${\bf f}$ in the model \eqref{stoch_model}. Thus this technique can be used to characterise the properties of only the stochastic
part $\varepsilon(t)$ of the model \eqref{stoch_model}.

In this technique, the scaling exponent $\alpha$ is a measure of the ratio of the logarithm of the {\it fluctuations} or vertical height of the graph to the
logarithm of the horizontal width of a chosen time window over which that vertical height is measured. The scaling exponent is calculated as the
slope of a log-log graph of a range of different horizontal time window sizes against, the vertical height of the signal in those time windows.
Mathematically, for self-similar signals with positive scaling exponent $\alpha$ the self-similarity property of the graph of the signal $s_n$
should hold on all time scales, but we are limited by the finite amplitude range of physical measurements to a certain maximum scale. Thus the signal
is first integrated in order to create a new stochastic process that exhibits self-similarity over a large range of time scales (then, for example, a purely independent,
stochastic process will result in a self-similar random walk).

First, the time series $s_n$ is integrated:
\begin{equation}
y_n = \sum_{j = 1}^n s_j ,
\end{equation}
for $n = 1, 2 \ldots N$, where $N$ is the number of samples in the signal $s_n$. Then, $y_n$ is divided into windows of length $L$ samples.
A least-squares straight line {\it local trend} is calculated by analytically minimising the squared error $E^2$ over the slope and intercept parameters $a$ and $b$:
\begin{equation}
\label{functional}
\mathop {\arg \min }\limits_{a,b} E^2  = \sum_{n = 1}^L \left( y_n - an - b \right)^2.
\end{equation}

Next, the root-mean-square deviation from the trend, the fluctuation, is calculated over every window at every time scale:
\begin{equation}
\label{rms-trend}
F( L ) = \left[ \frac{1}{L}\sum_{n = 1}^L \left( y_n - an - b \right)^2 \right]^{\frac{1}{2}}.
\end{equation}
This process is repeated over the whole signal at a range of different window sizes $L$, and a log-log graph of $L$ against $F(L)$ is constructed. A straight line
on this graph indicates self-similarity expressed as $F(L) \propto L^{\alpha}$. The scaling exponent $\alpha$ is calculated as the slope of a straight line fit to the log-
log graph of $L$ against $F(L)$ using least-squares as above. For a more in-depth presentation and discussion of self-similarity in signals in general, and further
information about DFA, please see Kantz, Hu \cite{kantz, hu}.

We are assuming that the signal, as the measured output of the new model, represents a combination of deterministic and stochastic dynamics.
The deterministic part of the dynamics, dictated by the function ${\bf f}$ in
equation \eqref{stoch_model} will result in slower changes in the signal $s_n$ taking place over a relatively long time scale.
Similarly, the stochastic fluctuations in the signal indicated changes taking place over a much shorter time scale. Since the goal of DFA is to analyse the faster changing,
stochastic properties of the signal, only a limited range of window sizes is investigated, over which the stochastic component of the signal exhibits
self-similarity as indicated by a straight-line on the log-log graph of window size against fluctuation.
As an example, Type III would include some speech signals that are actually chaotic, where the chaos is due to slow,
nonlinear dynamics in the vocal fold tissue and airflow, the characteristic time of this nonlinear oscillation will be much
longer than the window sizes over which the scaling exponent is estimated. Thus, the nature of the chaotic oscillation will
not affect the scaling exponent, which will respond only to any random fluctuations occurring on a much shorter time scale.

The resulting scaling exponent can assume any number on the real line. However, it would be more convenient to represent this scaling exponent
on a finite sliding scale from zero to one, as we have done for the RPDE measure. Thus we need a mapping function $g:\mathbb{R} \to [0, 1]$.
One such function finding common use in statistical and pattern recognition applications is the {\it logistic function}
$g(x) = {\left(1+ \exp(-x)\right)}^{-1}$ \cite{bishop-95}, so that the {\it normalised scaling exponent} is:
\begin{equation}
\alpha_{\rm norm} = \frac{1}{1+\exp\left( -\alpha \right)}.
\end{equation}
Therefore, each sound will lie somewhere between the extremes of zero and one on this scale, according to the self-similarity properties of the stochastic
part of the dynamics. As will be shown later, speech sounds for which $\alpha_{\rm norm}$ is closer to one are characteristic of general voice disorder.

\subsection*{Application to Examples of Normal and Disordered Voices}
In this section we will apply these two new measures to some examples of normal and disordered voices, as a limited demonstration of
characterising the extent of aperiodicity and breathiness in these signals.

Figure 4 shows the RPDE value $H_{\rm norm}$ calculated on the same two speech signals from the Kay database as shown in
figure 1. Note that the second, disordered example is of Type III and shows significantly irregular vibration, which is detected
by a large increase in $H_{\rm norm}$.

Similarly, figure 5 shows two more speech examples, one normal and one disordered from the same database and the corresponding
values of the scaling exponent $\alpha$ and $\alpha_{\rm norm}$. In these cases, the disordered example is extremely ``breathy'', and this turbulent noise is
detected by an increase in the scaling exponent.

\subsection*{Quadratic Discriminant Analysis}
In order to test the effectiveness of these two measures in practice, one approach is to set up a {\it classification task} to separate
normal control subjects from disordered examples using these measures alone. Here we choose one of the simplest approaches,
{\it quadratic discriminant analysis (QDA)}, which allows separation by modelling the data conditional upon each class,
here the normal (class $C_1$) and disordered (class $C_2$) cases, using joint Gaussian
probability density functions \cite{bishop-95}. For a $J \times K$ data matrix ${\bf v} = v_{jk}$ of observations consisting of the measures $j = 1,2$
for RPDE and DFA respectively, and all subjects $k$, these likelihood densities are parameterised by the mean and covariance matrices of the data set:
\begin{equation}
{\bm{\mu}} = E\left[ {\bf v} \right], \;\;{\bf C} = E\left[ {\left( {{\bf v} - {\bm{\mu}}} \right)\left( {{\bf v} - {\bm{\mu}}} \right)^T } \right],
\end{equation}
where $E$ is the expectation operator, and ${\bm{\mu}}$ is the mean vector formed from the means of each row of ${\bf v}$. The class likelihoods are:
\begin{equation}
f_C({\bf w}\left.\right| C_i) = \left(2 \pi\right)^{-J/2} \left| {\bf C}_i \right|^{-1/2}
\exp\left[ -\frac{1}{2} \left( {\bf w} - {\bm{\mu}_i} \right)^T {\bf C}_i^{-1} \left( {\bf w} - \bm{\mu}_i \right)\right],
\end{equation}
for classes $i = 1, 2$ and an arbitrary observation vector ${\bf w}$. It can be shown that, given these Gaussian class models,
the maximum likelihood regions of
the observation space $\mathbb{R}^J$ are separated by a {\it decision boundary} which is a (hyper-)conic section calculated from the difference of log-likelihoods
for each
class,
which is the unique set of points where each class is equally likely \cite{bishop-95}. The maximum likelihood classification problem is then solved using the
decision rule that $l({\bf w}) \ge 0$ assigns ${\bf w}$ to class $C_1$, and $l({\bf w}) < 0$ assigns it to class $C_2$, where:
\begin{gather}
l\left( {\bf w} \right) = -\frac{1}{2}{\bf w}^T {\bf A}_2 {\bf w} + {\bf A}_1 {\bf w} + A_0, \\
{\bf A}_2 = {\bf C}_1^{-1} - {\bf C}_2^{-1}, {\bf A}_1 = {\bm{\mu}}_1^T {\bf C}_1^{-1} - {\bm{\mu}}_2^T {\bf C}_2^{-1}, \\
A_0 = -\frac{1}{2} \ln \left| {\bf C}_1 \right | + \frac{1}{2} \ln \left| {\bf C}_2 \right | - \frac{1}{2}{\bm{\mu}}_1^T {\bf C}_1^{-1}{\bm{\mu}}_1 + \frac{1}{2}{
\bm{
\mu}}_2^T {
\bf C}_2^{-1}{\bm{\mu}}_2.
\end{gather}

In order to avoid the problem of overfitting, where the particular chosen separation model shows good performance on the training data but
fails to {\it generalise} well to new, unseen data, the classifier results require validation.

In this paper, we make use of {\it bootstrap resampling} for validation \cite{bishop-95}. In the bootstrap approach,
the classifier is trained on $K$ cases selected at random with replacement from the original data set of $K$ cases. This trial resampling processes is repeated
many times and the mean classification parameters $E\left[{\bf A}_2\right],E\left[{\bf A}_1\right],E\left[A_0\right]$ are selected as the parameters that would achieve
the best performance on entirely novel data sets.

Bootstrap training of the classifier involves calculating $H_{\rm norm}^k$ and $\alpha_{\rm norm}^k$ (the observations) for each speech sample $k$ in the database,
(where the superscript $k$ denotes the measure for the $k$-th subject). Then, $K$ random selections of these values with replacement
$H'^k_{\rm norm}$ and $\alpha'^k_{\rm norm}$ form the entries of the vector $v_{1k} = H'^k_{\rm norm}$ and $v_{2k} = \alpha'^k_{\rm norm}$
Then the mean vectors for each class $\bm{\mu}_1$ and $\bm{\mu}_2$ and covariance matrices ${\bf C}_1,{\bf C}_2$ for the whole selection are calculated.
Next, for each subject, the decision function is evaluated:
\begin{equation}
l({\bf w}_k) = l ([H_{\rm norm}^k, \alpha_{\rm norm}^k]^T).
\end{equation}
Subsequently, applying the decision rule assigns the subject $k$ into either normal or disordered classes. Then the performance of the classifier can be evaluated
in terms of percentage of true positives (when a disordered subject is correctly assigned to the disordered class $C_1$) and true negatives (when a normal subject
is correctly assigned to the normal class $C_2$). The overall performance is the percentage of correctly classified subjects, in both classes. This bootstrap
trial process of creating random selections of the measures, calculating the class mean vectors and covariance matrices, and then evaluating the decision function
on all the measures to obtain the classification performance is repeated many times. Assuming that the performance percentages are normally distributed, then
the 95\% confidence interval of the classification performance percentages can be calculated. The best classification boundary can then be taken as the
mean boundary overall all the trials.

Efficient implementations of the algorithms described in this paper written in C with Matlab MEX
interface accompany this paper: close returns [see Additional files 1 and 2] and
detrended fluctuation analysis [see Additional files 3, 4 and 5].

\subsection*{Algorithms for Classical Techniques}
For the purposes of comparison, we calculate the classical measures of jitter, shimmer and HNR (Noise-to-Harmonics Ratio) \cite{baken}.
There are many available algorithms for calculating this quantity, in this study we make use of the algorithms supplied in the software package
Praat \cite{praat}. These measures are based on an autocorrelation method for determining the pitch period (see Boersma \cite{boersma} for a detailed
description of the method).

We also use the methods described in Michaelis \cite{michaelis}. This first requires calculating the measures EPQ (Energy Perturbation Quotient),
PPQ (Pitch Perturbation Quotient), GNE (Glottal to Noise Excitation Ratio) and
the mean correlation coefficient between successive cycles, measures which require the estimation of the pitch period
using the waveform matching algorithm (see Titze \cite{titze-93} for a detailed description of this algorithm). The EPQ, PPQ, GNE and
correlation coefficients are calculated over successive overlapping ``frames'' of the speech signal. Each frame starts at a multiple of 0.26 seconds, and is
0.5 seconds long. For each frame, the EPQ, PPQ, GNE and correlation coefficients are combined into a pair of component measures,
called Irregularity and Noise. We use the average of the Irregularity and Noise components over all these frames \cite{michaelis}.

\subsection*{Classification Test Data}
\label{section_data}
This study makes use of the Kay Elemetrics disordered voice database (KayPENTAX Model 4337, New Jersey, USA), which contains 707 examples of
disordered and normal voices from a wide variety of organic, neurological and traumatic voice disorders. This represents examples of all three types of disordered
voice speech signals (Types I, II and III). There are 53 control samples from normal subjects. Each speech sample in the database was recorded under
controlled, quiet acoustic conditions, and is on average around two seconds long, 16 bit uncompressed PCM audio. Some of the speech samples
were recorded at 50kHz and then downsampled with anti-aliasing to 25kHz. Used in this study are sustained vowel phonations, since this controls for
any significant nonstationarity due changes in the position of the articulators such as the tongue and lips in running speech, which would have an
adverse effect upon the analysis methods. For calculating the Irregularity and Noise components, the signals are resampled with anti-aliasing to 44.1kHz.

\section*{Results}
\label{section_results}
Figure 6 shows hoarseness diagrams after Michaelis \cite{michaelis} constructed using the speech data and RPDE and DFA measures, the derived irregularity and
noise components of Michaelis, along with the same diagrams using two other combinations of the three classical perturbation measures for direct comparison.
The three classical measures are jitter, shimmer and NHR (Noise-to-Harmonics Ratio) \cite{baken}. The normalised RPDE, DFA scaling exponents and derived
irregularity and noise components are calculated for each of the $K = 707$ speech signals. Where the traditional perturbation algorithms did not fail to produce
a result, the traditional perturbation values were calculated
for a smaller subset of the subjects, see \cite{baken} for details of these traditional algorithms. Also shown in figure 6 is the result of the classification task applied to
the dataset; the best classification boundary is calculated using bootstrap resampling over 1000 trials. Table 1 summarises all the classification performance
results for the classification tasks on the hoarseness diagrams of figure 6.
The RPDE parameters were the same as for figure 3, and the DFA parameters were the same as for figure 5.

\section*{Discussion}
\label{section_discuss}
As shown in table 1, of all the combinations of the new and traditional measures, and derived irregularity and noise components, the highest overall correct
classification performance of $91.8 \pm 2.0\%$ is achieved by the RPDE/DFA pair. The combination of jitter and shimmer leads to the next highest performance.
These results confirm that, compared under the same, simple classifier approach, the new nonlinear measures are more accurate on average than traditional measures or
the derived irregularity and noise components. We will now discuss particular aspects of these results in comparison with traditional perturbation measures.

\subsection*{Feature Dimensionality}
The {\it curse of dimensionality} afflicts all challenging data analysis problems \cite{bishop-95}. In pattern analysis tasks such as automated normal/disordered
separation, increasing the size of the feature vector (in this case, the number of measures $J$ in the classifier vector ${\bf v}$) does not necessarily increase the performance of the
classifier in general. This is because the volume of the {\it feature space} (the space spanned by the possible values of the measures) grows exponentially with the
number of features. Therefore, the limited number of examples available to train the classifier occupy an increasingly small volume in the feature space, providing a poor representation
of
the mapping from features to classes that the classifier must learn \cite{bishop-95}. Therefore the new measures help to mitigate this problem of dimensionality, since only these two
new measures are required to obtain good separation performance. By comparison, we need to calculate four different measures in order obtain the irregularity and
noise components \cite{michaelis}.

\subsection*{Feature Redundancy -- Information Content}
It is also important to use as few features as possible because in practice, increasing the number of features causes excessive data to be generated that may well contain
redundant (overlapping) information. The actual, useful information contained in these vectors has a much smaller dimensionality. For clinical purposes, it is important that only this
useful data is
presented. This effect of redundant information for the traditional measures can be clearly seen in figure 6, where combinations of pairs of (the logarithms of) these measures
are seen to cluster around a line or curve in the feature space, indicating high positive correlation between these measures. Traditional measures occupy an effectively
one-dimensional object in this two-dimensional space. The irregularity and noise components occupy more of the area of the feature space than traditional measures, and the new
measures are spread evenly over the same space.

\subsection*{Arbitrary Parameters -- Reproducibility}
Minimising the number of arbitrary parameters used to calculate these measures is necessary to avoid selecting an excessively specialised set of parameters
that leads, for example, to good normal/disordered separation on a particular data set but does not generalise well to new data.

Many parameters are required for the algorithms used in calculating traditional perturbation measures \cite{michaelis, boyanov, alonso}. For example, the waveform
matching algorithm \cite{baken} requires the definition of rough markers, upper and lower pitch period limits, low-pass filter cutoff frequencies, bandwidth and order selection parameters,
and the number of
pitch periods for averaging should these pitch period limits be exceeded \cite{titze-93}. Similarly, in just one of the noise measures (glottal-to-noise excitation ratio) used in
Michaelis \cite{michaelis}, we can determine explicitly at least
four parameters relating to linear prediction order, bandpass filter number, order, cutoff selection, and time lag range parameters.
There are two additional parameters for the length and starting sample of the part of the signal selected for analysis.

Our new measures require only five arbitrary parameters that must be chosen in advance: the length of the speech signal $N$, the maximum recurrence time
$T_{\rm max}$, and the lower value, upper value and increment of the DFA interval lengths $L$. We have also shown, using analytical results, that we can calibrate out the
dependence upon the state space close recurrence radius $r$, the time-delay reconstruction dimension $d$ and the reconstruction delay $\tau$.

\subsection*{Interpretation of Results}
We have found, in agreement with Titze \cite{titze-95} and Carding \cite{carding}, that perturbation measures cannot be obtained for all the speech sounds produced
by subjects (see table 1). This limits the clinical usefulness of these traditional measures. By contrast, the new measures presented in this chapter do not
suffer from this limitation and are capable of measuring, by design, all types of speech signals.

Taking into account the classification performance achievable using a simple classifier, the number of these measures that need to be combined to achieve effective
normal/disordered separation, the number of arbitrary parameters used to calculate the measures, and the independence of these measures, traditional approaches and derived irregularity
and noise
components are seen to be considerably more complex than the new measures developed in this paper. The results of the classification comparison with traditional
measures and the irregularity and noise components suggest that, in order to reach the classification performance
of the new measures, we will either need much more complex classifiers, or need to combine many more classical features together \cite{boyanov, godino-04, alonso}.
It is therefore not clear that traditional approaches capture the {\it essential biomechanical differences} between normal and disordered voices in the most
parsimonious way, and an excessively complicated relationship exists therefore between the values of these measures and extent of the voice disorder.
As a final comment, we note that the classical perturbation measures were, for the majority of signals, able to produce a result regardless of the type of the signal. This is consistent
with the findings of other studies \cite{michaelis}, where for Type II/III and random noise signals, the correct interpretation of these measures breaks down.
Therefore, although it is no longer possible in these cases to assign a coherent meaning to the results produced by these measures, this does not {\it per se} mean that there is
not some as yet unknown connection between disorder and the these measures. For this reason, we do not discard the results of these measures for Type II/III and random
cases.

\subsection*{Limitations of the New Measures}
There are certain limitations to the new measures in clinical practice. These measures rely upon sustained vowel phonation, and sometimes
subjects experience difficulty in producing such sounds, which limits the applicability. Also, at the beginning of a sustained vowel phonation, the voice of many
subjects may require some time to settle into a more stable vibration. As such, discarding the beginning of the phonation is usually a prerequisite (but
this does not adversely affect the applicability of these methods). Nonetheless, the extent of breathiness in speech is not usually affected
in this way. In practice we require that the subject maintains a constant distance from the microphone when producing speech sounds;
this can be achieved, for example, with the use of head-mounted microphones. We note that these limitations also apply to existing measures.

\subsection*{Possible Improvements and Extensions}
There are several improvements that could be made to these new measures. Firstly, every arbitrary parameter introduces extra variability
that affects the reliability of the results. Much as it has been possible to calibrate out the dependence upon the RPDE parameters
using analytical results, a theoretical study of the DFA interval lengths based upon typical sustained phonation recurrence
periods could reveal values that would be found for all possible speech signals. These would be related to the sampling time $\Delta t$.
The particular choice of normalisation function $g$ for the scaling exponent might affect the classification performance, and better
knowledge of the possible range of $\alpha$ values using theoretical studies of the DFA algorithm would be useful for this.
It should also be possible to increase the recurrence time precision of the RPDE analysis by interpolating the state space
orbits around the times of close recurrence $n_0, n_1$. It should then be possible to achieve the same high-resolution as waveform matching
techniques which would make RPDE competitive for the detailed analysis of Type I periodic sounds.

\section*{Conclusions}
\label{section_summary}
In this study, in order to directly characterise the two main biophysical factors of disordered voices: increased nonlinear, complex aperiodicity and non-Gaussian,
aeroacoustic breath noise, we have introduced recurrence and scaling analysis methods. We introduced a new, combined nonlinear/stochastic signal
model of speech production that is capable, in principle, of producing the wide variation in behaviour of normal and disordered voice examples.
To exploit the output of this model in practice, and hence all types of normal and disordered voices, we explored the use of two nonlinear measures:
the recurrence period density entropy and detrended fluctuation analysis.

Our results show that, when the assumptions of the model hold under experimental conditions (in that the speech examples are sustained vowels recorded
under quiet acoustic conditions), we can directly characterise the two main factors of aperiodicity and breath noise in disordered voices and thus construct
a ``hoarseness'' diagram showing the extent of normality/disorder from a speech signal. The results also show that on average, over all bootstrap resampling
trials, these two measures alone are capable of distinguishing normal subjects from subjects with all types of voice disorder, with better classification performance
than existing measures.

Furthermore, taking into account the number of arbitrary parameters in algorithms for calculating existing perturbation measures, and the number of these
existing measures that need to be combined to perform normal/disordered separation,
we have shown that existing approaches are considerably more complex.

We conclude that the nonlinearity and non-Gaussianity of the biophysics of speech production can be exploited in the design of signal analysis methods
and screening systems that are better able characterise the wide variety of biophysical changes arising from voice disease and disorder. This is because
ultimately the biophysics of speech production generate the widely varying phenomenology.

\section*{Appendix -- Mathematical Proofs}

\subsection*{Periodic Recurrence Probability Density}
We consider the purely deterministic case, i.e. when the model of equation \eqref{stoch_model} has no
forcing term ${\bf \varepsilon} (t)$. Thus the measured time series is purely deterministic
and points in the time series follow each other in an exactly prescribed sequence. When the measured,
trajectory ${\bf s}_n$ is a purely periodic orbit of finite period $k$ steps, there is an infinite sequence of
points $\{ {\bf r}_n \},  n \in {\bf Z}$ in the reconstructed state space with ${\bf r}_n  = {\bf r}_{n + k}$,
and ${\bf r}_n \ne {\bf r}_{n + j}$ for $0 < j < k$.

Picking any point ${\bf s}$ in the reconstructed state-space, there are two cases to
consider. In the first case, if ${\bf s} = {\bf r}_n$ for some $n$, then ${\bf s}$ is not the same as any other points in the periodic orbit
except for ${\bf r}_{n + k}$, so that the trajectory returns
with certainty for the first time to this point after $k$ time steps. This certainty, with the requirement the that probability
of first recurrence is normalised for $T = 1, 2 \ldots $ implies that:
\begin{equation}
P_{\bf s} \left( {T = r} \right) = \left\{ \begin{array}{l}
 1\quad {\rm if}\;r = k \\
 0\quad {\rm otherwise} \\
 \end{array} \right. .
\end{equation}

In the second case when ${\bf s} \ne {\bf r}_n$ for any $n$, the trajectory never intersects the point so that there are also
never any first returns to this point. All the points in the reconstructed space form a disjoint partition of the whole
space. Thus the probability of recurrence to the whole space is the sum of the probability of recurrence to each point in the
space separately, appropriately weighted to satisfy the requirement that the probability of first recurrence to the whole space is normalised
However, only the $k$ distinct points of the periodic orbit contribute to the total probability of first recurrence to the whole space. Therefore,
the probability of first recurrence is:
\begin{equation}
P(T) = \frac{1}{k} \sum\limits_{i = 0}^{k - 1} {P_{{\bf r}_i } \left( {T = r} \right)}  = \left\{ \begin{array}{l}
 1\quad {\rm if}\;r = k \\
 0\quad {\rm otherwise} \\
 \end{array} \right. .
\end{equation}

\subsection*{Uniform i.i.d. Stochastic Recurrence Probability Density}
Consider the purely stochastic case when the nonlinear term ${\bf f}$ is in equation \eqref{stoch_model} is zero and the stochastic forcing
term is an i.i.d. random vector. Then the measured trajectory ${\bf s}_n$ is also a stochastic, i.i.d. random vector.
Since all the time series are normalised to the range $[-1, 1]$ then each member of the measurement takes on a value from this range.
Then the trajectories ${\bf s}_n$ occupy the reconstructed state-space which is the region $[-1, 1]^m$,
and each co-ordinate $s_n$ is i.i.d. uniform. We form a equal-sized partition of this space into $N^m$ (hyper)-cubes,
denoting each cubical region $R$. The length of the side of each cube $R$ is $\Delta s = 2/N$. Then the probability of finding the trajectory
in this cube is $P_R = \Delta s^m / 2^m$. Since the co-ordinates $s_n$ are uniform i.i.d.,
then the probability of first recurrence of time $T$ to this region $R$ is geometric \cite{altmann}:
\begin{equation}
P_R \left( T \right) = P_R \left[ {1 - P_R } \right]^{T - 1}  = \frac{{\Delta s^m }}{{2^m }}\left[ {1 - \frac{{\Delta s^m }}{{2^m }}} \right]^{T - 1} .
\end{equation}

This is properly normalised for $T = 1, 2 \ldots$. However, we require the probability of
first recurrence to all possible cubes. The cubes are a disjoint partition of the total reconstruction space $[-1, 1]^m$.
Thus the probability of recurrence to the whole space is the sum of the probability of recurrence to each cube separately, appropriately
weighted to satisfy the requirement that the probability of recurrence to the whole space is normalised. Since the probability of
first recurrence to each cube $R$, $P_R(T)$ is the same, the probability of recurrence to all cubes is:
\begin{align}
P(T) &= \sum_{i = 1}^{N^m} \frac{\Delta s^m }{2^m } P_R ( T )  = N^m \frac{\Delta s^m }{2^m } P_R ( T )\\
  &= \frac{2^m }{\Delta s^m } \frac{\Delta s^m }{2^m} P_R \left[ 1 - P_R \right]^{T - 1}  = \frac{\Delta s^m }{2^m }\left[ 1 - \frac{\Delta s^m }{2^m} \right]^{T
  - 1} .
\end{align}

For small cube side lengths $\Delta s$ and close returns algorithm radius $r$, the first recurrence probability determined by the close
returns algorithm is then:
\begin{equation}
P( T ) = \frac{\Delta s^m }{2^m }\left[ 1 - \frac{\Delta s^m }{2^m} \right]^{T - 1} \approx \frac{r^m }{2^m }\left[ 1 - \frac{r^m }{2^m } \right]^{T - 1}.
\end{equation}

Similarly, for small close returns radius $r$ and/or for large embedding dimensions $m$, $1-r^m/2^m \approx 1$ so that:
\begin{equation}
P( T) \approx \frac{r^m }{2^m }.
\end{equation}
Note that for fixed $m$ and $r$ this expression is constant. Since the close returns algorithm can only measure recurrence periods over a limited
range $ 1 \le T \le T_{\rm max}$, and we normalise the recurrence histogram $R(T)$ over this range of $T$, then
the probability of first recurrence is the uniform density:
\begin{equation}
P(T) \approx \frac{1}{T_{\rm max}},
\end{equation}
which is proportional to the expression $r^m / 2^m$ above. Thus, up to a scale factor, for a uniform i.i.d. stochastic signal, the recurrence
probability density is uniform.

\section*{Authors' Contributions}
MAL lead the conceptual design of the study, developed the mathematical methods, wrote the software and data analysis tools,
and prepared and analysed the data. PEM participated in the conceptual design of the study and the mathematical
methods. SJR participated in the discriminant analysis of the data. DAEC participated in the data preparation.
IMM contributed to the development of the mathematical methods. All authors read and approved the manuscript.

\section*{Acknowledgements}
  \ifthenelse{\boolean{publ}}{\small}{}
Max Little acknowledges the financial support of the Engineering and Physical Sciences Research Council, UK, and wishes to thank
Prof. Gesine Reinert (Department of Statistics, Oxford University, UK), Mr Martin Burton (Radcliffe Infirmary, Oxford, UK)
and Prof. Adrian Fourcin (University College London, London, UK) for valuable discussions and comments on early drafts of this paper.

{\ifthenelse{\boolean{publ}}{\footnotesize}{\small}
 \bibliographystyle{bmc_article}  
  \bibliography{bmeo_rec_frac} }     

\ifthenelse{\boolean{publ}}{\end{multicols}}{}

\section*{Figures}

\begin{figure}[ht]
\centering
\includegraphics{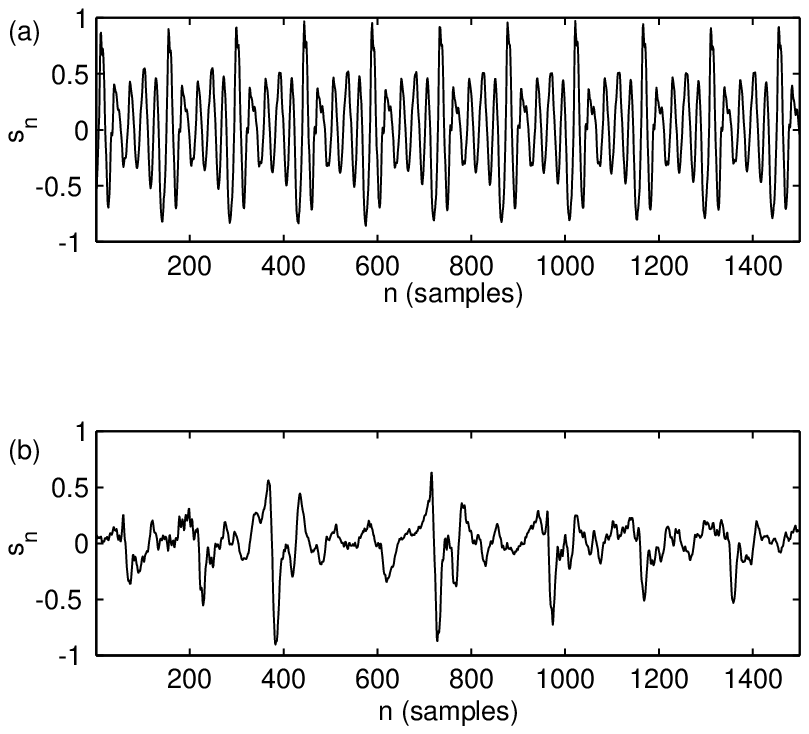}
\caption[Selected normal and disordered speech signal examples]{Discrete-time signals from (a) one normal (JMC1NAL) and (b) one disordered (JXS01AN) speech signal from the Kay
Elemetrics database. For clarity only a small section is shown (1500 samples).}
\label{fig-1}
\end{figure}

\begin{figure}[ht]
\centering
\includegraphics{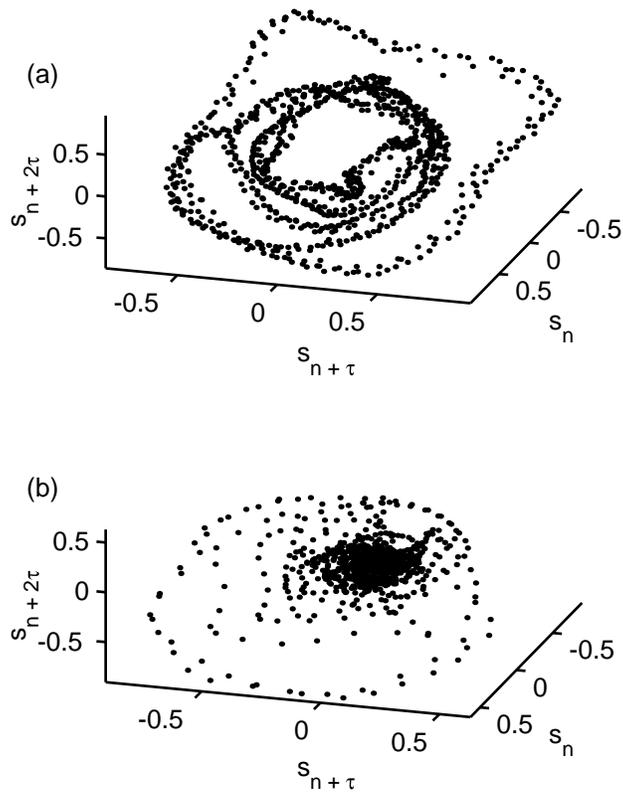}
\caption[Selected time-delay embedded speech signals]{Time-delay embedded discrete-time signals from (a) one normal (JMC1NAL) and (b) one disordered (JXS01AN) speech signal from
the Kay Elemetrics database. For clarity only a small section is shown (1500 samples). The embedding dimension is $m=3$ and the time delay is $\tau = 7$ samples.}
\label{fig-2}
\end{figure}

\begin{figure}[ht]
\centering
\includegraphics{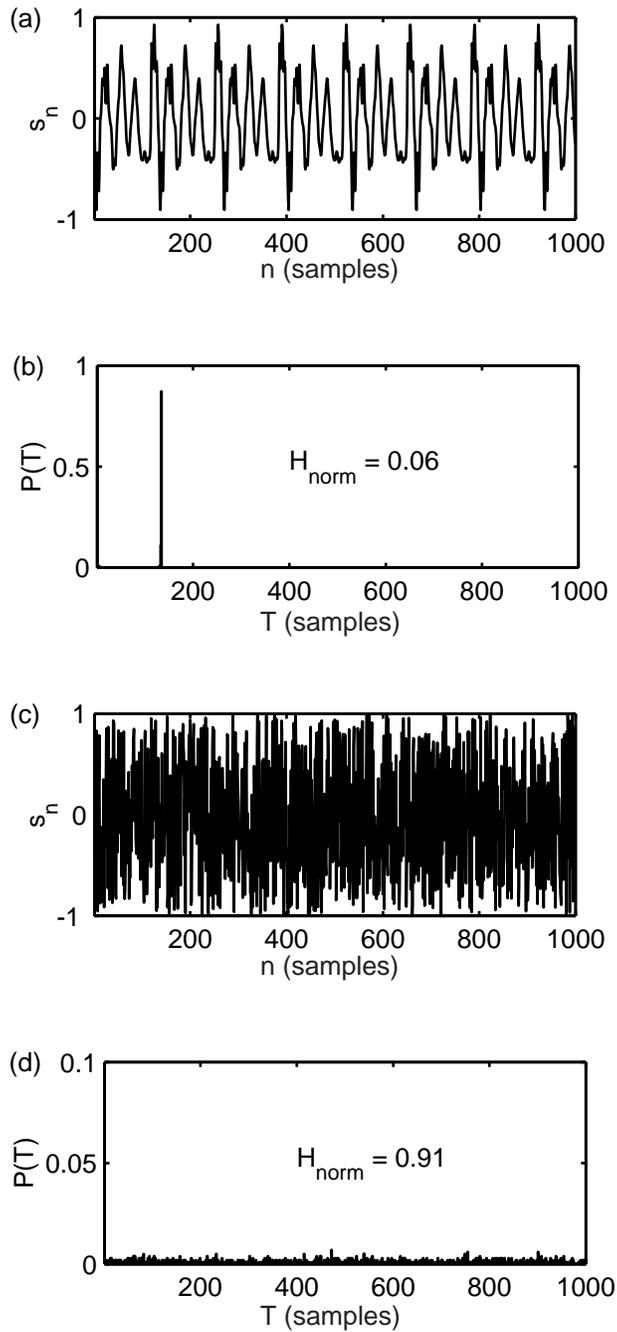}
\caption[State-space recurrence analysis for a periodic signal]{Demonstration of results of time-delayed state-space recurrence analysis applied to a perfectly periodic signal (a) created by
taking a single cycle (period $k = 134$ samples) from a speech signal and repeating it end-to-end many times. The signal was normalised to the range $[-1, 1]$. (b) All values of $P(T)$
are zero except for $P(133) = 0.1354$ and $P(134) = 0.8646$ so that $P(T)$ is properly normalised. This analysis is also applied to (c) a synthesised, uniform i.i.d. random signal on
the range $[-1, 1]$, for which (d) the density $P(T)$ is fairly uniform. For clarity only a small section of the time series (1000 samples) and the recurrence time (1000 samples) is
shown. Here, $T_{\rm max} = 1000$. The length of both signals was 18088 samples. The optimal values of the recurrence analysis parameters were found at $r = 0.12$, $m=4$ and
$\tau = 35$.}
\label{fig-3}
\end{figure}

\begin{figure}[ht]
\centering
\includegraphics{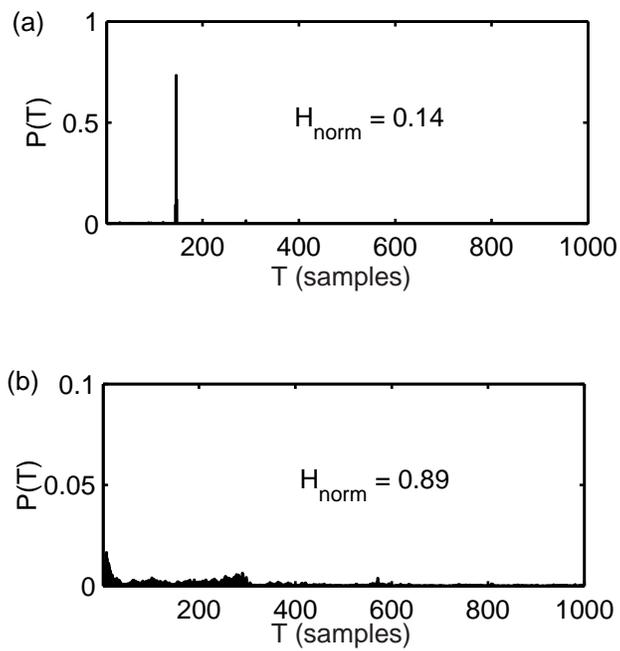}
\caption[RPDE analysis results]{Results of RPDE analysis carried out on the two example speech signals from the Kay database as shown in figure 1. (a) Normal voice (JMC1NAL), (b)
disordered voice (JXS01AN). The values of the recurrence analysis parameters were the same as those in the analysis of figure 3. The normalised RPDE value $H_{\rm norm}$ is larger
for the disordered voice.}
\label{fig_4}
\end{figure}

\begin{figure}[ht]
\centering
\includegraphics{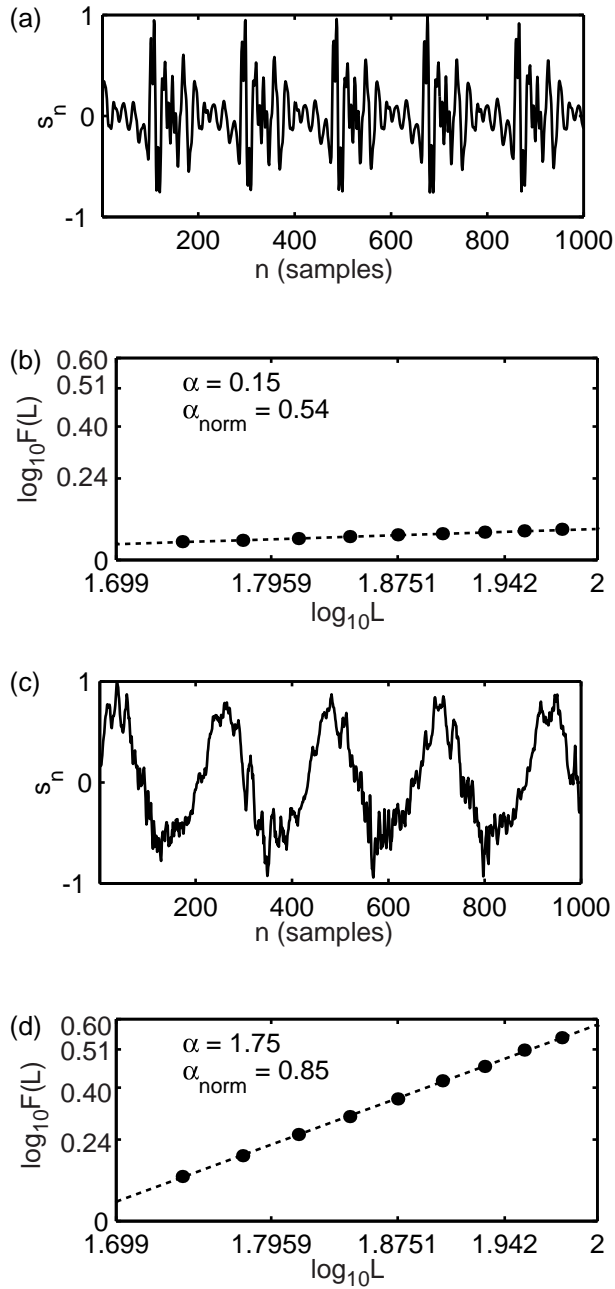}
\caption[DFA analysis results]{Results of scaling analysis carried out on two more example speech signals from the Kay database. (a) Normal voice (GPC1NAL) signal, (c) disordered
voice (RWR14AN). Discrete-time signals $s_n$ shown over a limited range of $n$ for clarity. (b) Logarithm of scaling window sizes $L$ against the logarithm of fluctuation size $F(L)$
for normal voice in (a). (d) Logarithm of scaling window sizes $L$ against the logarithm of fluctuation size $F(L)$ for disordered voice in (b). The values of $L$ ranged from $L = 50$ to
$L =  100$ in steps of five. In (b) and (d), the dotted line is the straight-line fit to the logarithms of the values of $L$ and $F(L)$ (black dots). The values of $\alpha$ and the normalised
version $\alpha_{\rm norm}$ show an increase for the disordered voice.}
\label{fig_5}
\end{figure}

\begin{figure}[ht]
\centering
\includegraphics{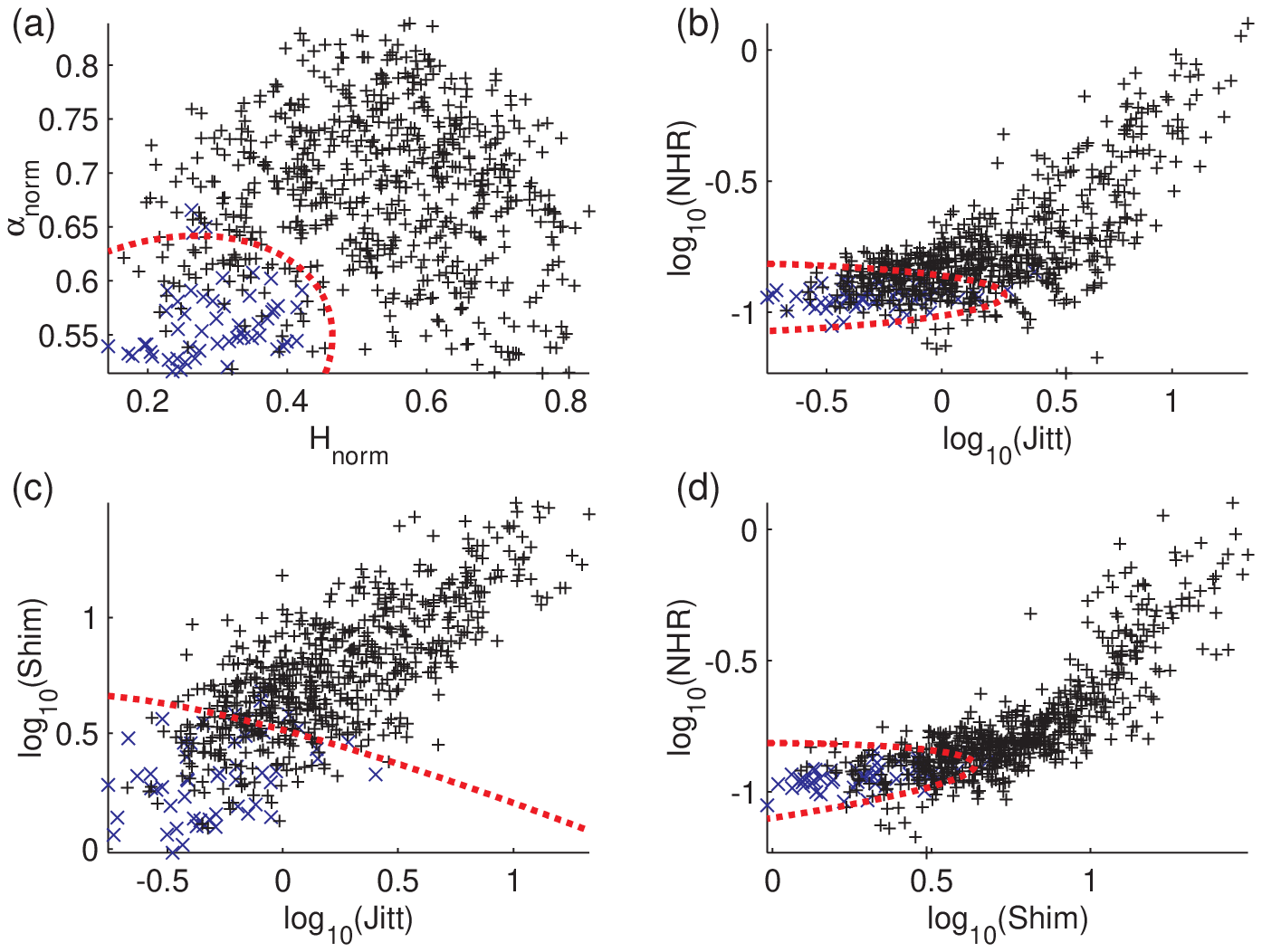}
\caption[``Hoarseness'' diagrams]{``Hoarseness'' diagrams illustrating graphically the distinction between normal (blue 'x' symbols) and disordered (black '+' symbols) on all speech
examples from the Kay Elemetrics dataset, for (a) the new measures return period density entropy (RPDE) (horizontal axis) and detrended fluctuation analysis (DFA) (vertical axis), (b)
for the irregularity (horizontal) and noise (vertical) components of Michaelis \cite{michaelis}, (c) for  classical perturbation measures jitter (horizontal) and noise-to-harmonics ratio (
NHR) (vertical) and (d) shimmer (horizontal) against NHR (vertical). The red dotted line shows the best normal/disordered classification task boundary over 1000 bootstrap trials using
quadratic discriminant analysis (QDA). The values of the RPDE and DFA analysis parameters were the same those in the analysis of figures 3 and 5 respectively. The logarithm of the
classical perturbation measures was used to improve the classification performance with QDA.}
\label{fig_6}
\end{figure}

\section*{Tables}
  \subsection*{Table 1 - Summary of disordered voice classification results}
   Summary of disordered voice classification task performance results, for several different combinations of the
   new measures, the derived irregularity (Irreg) and noise (Noise) components of Michaelis \cite{michaelis},
   and traditional perturbation measures, jitter (Jitt), shimmer (Shim) and noise-to-harmonics
   ratio (NHR). The RPDE parameters were the same as for figure 4, and the DFA parameters were
   the same as for figure 5. \par \mbox{}
    \par
    \mbox{\begin{tabular}{p{1.2in}p{1.0in}p{0.9in}p{0.9in}p{0.8in}}
    \hline
    Combination          & Subjects  & True Positive  & True Negative   & {\bf Overall} \\
    \hline
    RPDE/DFA           & 707 &             95.4$\pm$3.2\%  &  91.5$\pm$2.3\% &  {\bf 91.8$\pm$2.0\%} \\
    Jitt/Shim                & 685 &             86.9$\pm$6.9\%  &  81.0$\pm$4.7\% &  {\bf 81.4$\pm$3.9\%} \\
    Shim/NHR             & 684 &             91.4$\pm$5.9\%  &  79.8$\pm$4.7\% &  {\bf 80.7$\pm$4.0\%} \\
    Irreg/Noise             & 707 &            78.4$\pm$6.2\%   &  90.5$\pm$4.9\% &  {\bf 79.3$\pm$5.5\%} \\
    Jitt/NHR                & 684 &            93.2$\pm$7.4\%   &  75.0$\pm$5.5\% &  {\bf 76.4$\pm$4.8\%} \\
    \hline
    \end{tabular}
    }

\section*{Additional Files}
  \subsection*{Additional file 1 -- close\_ret.c, 6K}
    Close returns algorithm implemented in C with Matlab MEX interface. Standard ASCII text file format.

  \subsection*{Additional file 2 -- close\_ret.dll, 53K}
    Close returns algorithm compiled as a DLL for Windows. Standard Windows DLL format.

  \subsection*{Additional file 3 -- fastdfa\_core.c, 9K}
    Efficient implementation of the detrended fluctuation analysis (DFA) algorithm written
    in C with Matlab MEX interface. Core C code. Standard ASCII text file format.

  \subsection*{Additional file 4 -- fastdfa\_core.dll, 53K}
    DFA algorithm core compiled as a DLL for Windows. Standard Windows DLL format.

  \subsection*{Additional file 5 -- fastdfa.m, 2K}
    Matlab function wrapper for above DFA algorithm implementation. Standard ASCII Matlab script file format.
    Type 'help fastdfa.m' at the Matlab command prompt for usage instructions.

\end{bmcformat}
\end{document}